\documentclass[preprint,12pt]{elsarticle}

\usepackage{graphicx}
\usepackage{amssymb}

\usepackage{lineno}

\journal{Journal Name}

\begin{document}

\begin{frontmatter}


\title{Influence of cutoff dipole interaction radius and dilution on phase transition in kagome artificial spin ice}

 \author{Petr Andriushchenko\fnref{label1,label2}}
 \ead{pitandmind@gmail.com}

 \address[label1]{School of Natural Sciences, Far Eastern Federal University, Sukhanova 8, Vladivostok, Russian Federation}
 \address[label2]{Institute of Applied Mathematics, Far Eastern Branch, 
Russian Academy of Science,Radio 7, Vladivostok, Russian Federation}



\begin{abstract}
The investigations carried out in this paper show the influence of cutoff dipole interaction radius and dilution on the basic thermodynamic characteristics of the kagome artificial spin ice with free boundary conditions. The phase transition disappearance at the dipole interaction radius limitation was shown in our previous study by Shevchenko {\it et al.} [JETP {\bf 124}(6), 982–993 (2017)]. This work continues this study and answers the question of the coordination spheres numbers that must be taken into account, in order not to lose the basic collective phenomena. The answer is at least three coordination spheres with 14 nearest neighbors must be taken into account for kagome artificial spin ice. Restriction to a smaller cutoff radius leads to significant changes in the thermodynamic behavior of the main characteristics of the system. An increase of the interaction radius shifts the phase transition temperature to the low-temperature area. The effects of dilution on kagome artificial spin ice were also investigated. It is shown that the phase transition occurs up to $p_c\approx0.35$ dilution concentrations, that well coincides with the Sykes and Essam theory. Further dilution leads to the phase transition disappearance.
\end{abstract}


\end{frontmatter}

\linenumbers

\section{Introduction}
\label{S:1}

In recent years, spin ice has become a very researchable topic. Thanks to frustrations (or competing interactions) a new physics appears, for the description of which new researches are needed \cite{zvyagin2013new}. Such systems have very rich behavior, but they are difficult to study because of the atomistic scale and three-dimensional lattice \cite{gingras2011spin}. The researchers attempted to facilitate the task themselves, and try to transferred studies of effects caused by frustrations to more ``simple'' systems - the artificial spin ice \cite{wang2006artificial}. Artificial spin ice is an patterned array of nano-islands (macrospins) made of magnetic material disposed on a nonmagnetic substrate and formed into a specific geometry. As Wang wrote in his work ``artificial frustrated magnet opens the door to a new mode of research wherein a frustrated system can be designed rather than discovered'' \cite{wang2006artificial}. The main distinctive feature of artificial spin ice from spin ice is that due to its mesoscopic scale and two-dimensionality, both the state of each island and the state of the system as a whole can be investigated using contemporary technologies. The modern development of experimental approaches allows observing static magnetic states of the systems using MFM \cite{wang2006artificial,budrikis2012domain,tanaka2006magnetic,schumann2010charge, kaur2017magnetic}, it is even possible to observe dynamic magnetic states in real time with X-PEEM \cite{rougemaille2011artificial} and XMCD-PEEM \cite{canals2016fragmentation,gliga2017emergent,farhan2013exploring, kapaklis2014thermal} methods. Gartside {\it et al.} in recent work \cite{gartside2018realization} showed the ``topological defect-driven magnetic writing'' method with which one can set the direction of magnetization for each macrospin in the array of artificial spin ice. Thus, with the help of this method, it is now possible to obtain and investigate any magnetic state of such systems. 

In the square artificial spin ice, which was first investigated in \cite{wang2006artificial}, four spins are meeting in each lattice vertex. Because of the dipole-dipole interactions, it is advantageous for the neighboring spins to line up ``head-to-tail'', so the ground states configurations will obey to the so-called ``ice rule'' when two spins point into the vertex and other two spins point out. That is, if one denote the head of the magnetization vector of each islands for ``$+$'' and the tail for ``$-$'' and sum them at each vertex of the lattice, we get dumbbell magnetostatic charge model \cite{budrikis2012domain,castelnovo2008magnetic}. If any vertex has a non zero charge, then this means some excitation in this area. 

The kagome lattice is interesting primarily for its geometry. In contrast to square artificial spin ice, at each vertex (with the exception of hexagons at the edges) of each hexagon, three magnetic islands are meeting instead of four (see Fig. \ref{dipole_model}), so the ice rule cannot be fulfilled at all. Thus, the kagome lattice is more frustrated, and therefore if one studying the effects caused by frustrations, the state space of such a system is more interesting.

\section{Model}
\label{S:2}

\begin{figure}
	\begin{minipage}[h]{0.49\linewidth}
		\centering\includegraphics[width=1\linewidth]{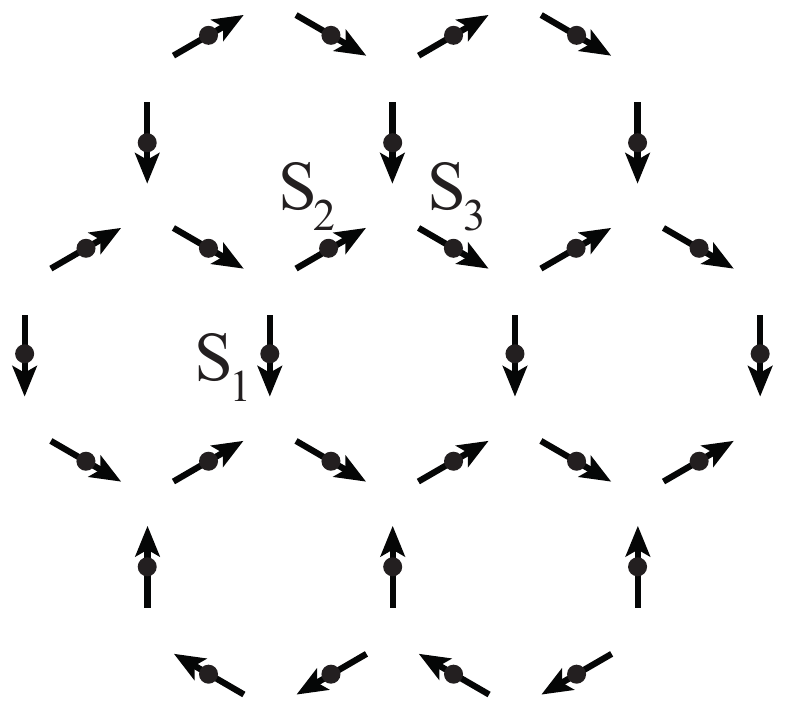} \\ a)
	\end{minipage}
	\hfill
	\begin{minipage}[h]{0.49\linewidth}
		\centering\includegraphics[width=1\linewidth]{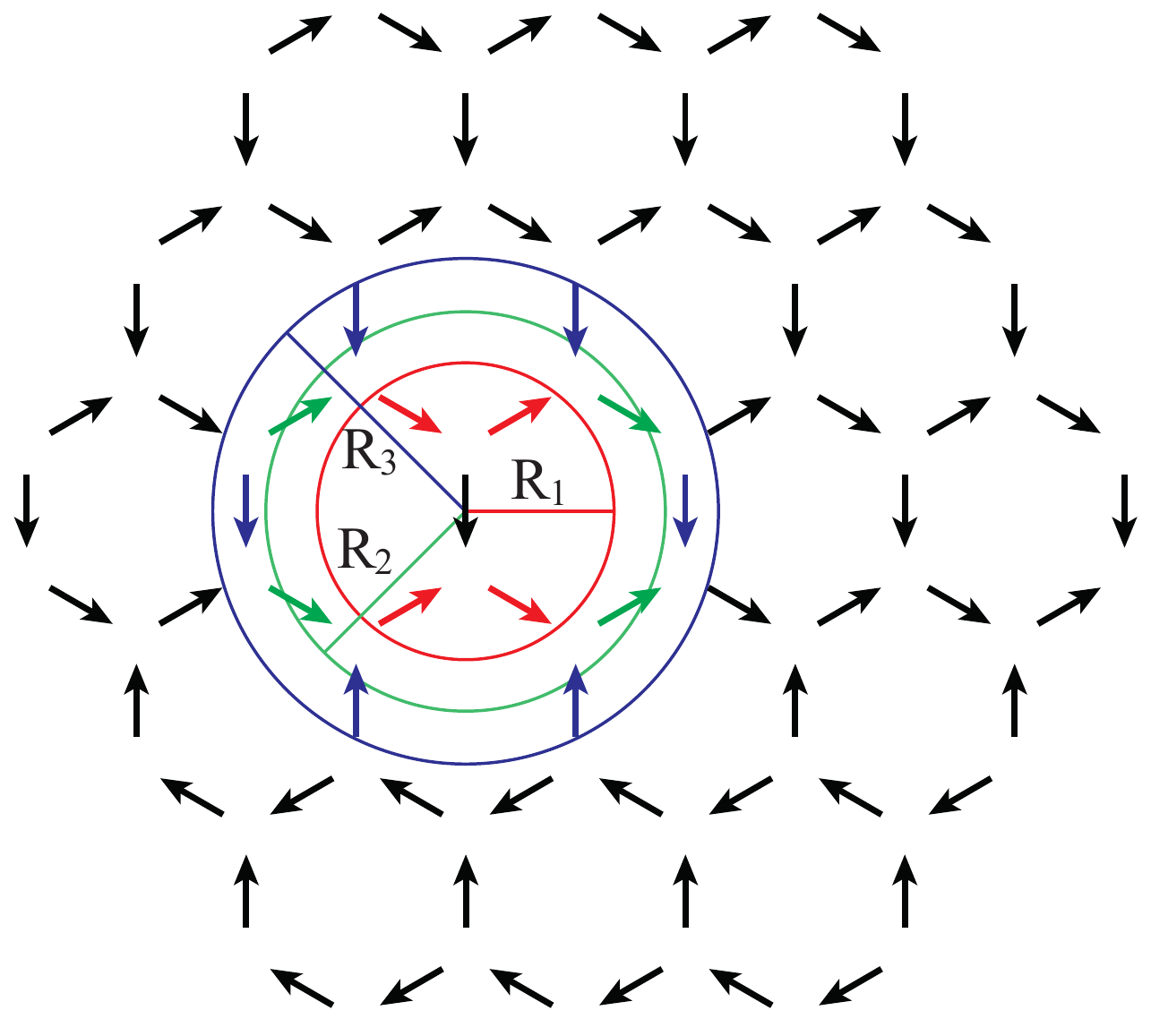} \\b)
	\end{minipage}
	\caption{a) Kagome artificial spin ice model with dipole-dipole interactions (hexagonal sample with 30 macrospins). Each macrospin represented by point dipole (denoted by black dots, in the remaining figures points will be omitted) with Ising-like magnetic moment (denoted by arrows) b) Three coordination spheres (cutoff radii) for the kagome artificial spin ice (hexagonal sample with 72 macrospins): $R_1$ (red), includes only 4 nearest neighbors, $R_2$ (green), includes 8 nearest neighbors and $R_3$ (blue), includes 14 nearest neighbors.}
	\label{dipole_model}
\end{figure}

Each magnetic island (macrospin) in artificial spin ice usually have a form anisotropy (for example length, width, and thickness of 220 nm $\times$ 80 nm $\times$ 25 nm in \cite{wang2006artificial} or 63 nm $\times$ 26 nm $\times$ 6 nm in \cite{anghinolfi2015thermodynamic} or even 1000 nm $\times$ 75 nm $\times$ 10 nm in \cite{gartside2018realization}), which limits the degree of freedom of the island magnetic moment and causes it to align along the long axis, at the same time remaining single-domain. In this work, the artificial spin ice model with dipole-dipole interaction described in the previous study \cite{shevchenko2017multicanonical, shevchenko2017effect} also in \cite{silva2012thermodynamics,greenberg2018disordered} is investigated. According to this model, each macrospin is represented by a point dipole, the magnetic moment of which have the Ising-like behavior. Detailed study of the usage of dipole interaction between two-dimensional magnetic particles was carried out by Politi and Pini in \cite{politi2002dipolar} and Leon and Pozo in \cite{leon2008using}. Thus, in this approach the kagome artificial spin ice model of hexagonal shape (for symmetry) with free boundary conditions is investigated (see Fig. \ref{dipole_model}a). Each macrospin represented by point dipole (denoted by dots) with Ising-like magnetic moment (denoted by arrows) which equal $\sigma_i \vec{m}_i$, where $\sigma_i=\pm1$ and $\vec{m}_i=\{0,-1\}$ for $S_1$, $\vec{m}_i=\{\sqrt{3/4},0.5\}$ for $S_2$ and $\vec{m}_i=\{\sqrt{3/4},-0.5\}$ for $S_3$ on the Fig.\ref{dipole_model}a example. The system size changing was occurred by successively adding layers to the hexagonal form to preserve the shape symmetry. The dipole-dipole interaction energy for two $ij$ spins was 

\begin{equation}
E^{ij}_{dip}=D \left(
\frac{
	({\vec m}_i{\vec m}_j)
}{
	\vert{\vec r}_{ij}\vert^3
}
-3
\frac{
	({\vec m}_i{\vec r}_{ij})
	({\vec m}_j{\vec r}_{ij})
}{
	\vert{\vec r}_{ij}\vert^5
}\right),
\label{dipint}
\end{equation}
where  
$D=\mu^2/a^3$ is a dimensional coefficient, $\mu$ is the total magnetic moment of the island, and $a$ is the lattice parameter \cite{shevchenko2017multicanonical,shevchenko2017effect}.

\begin{figure}
\begin{minipage}[h]{0.49\linewidth}
	\center{\includegraphics[width=1\linewidth]{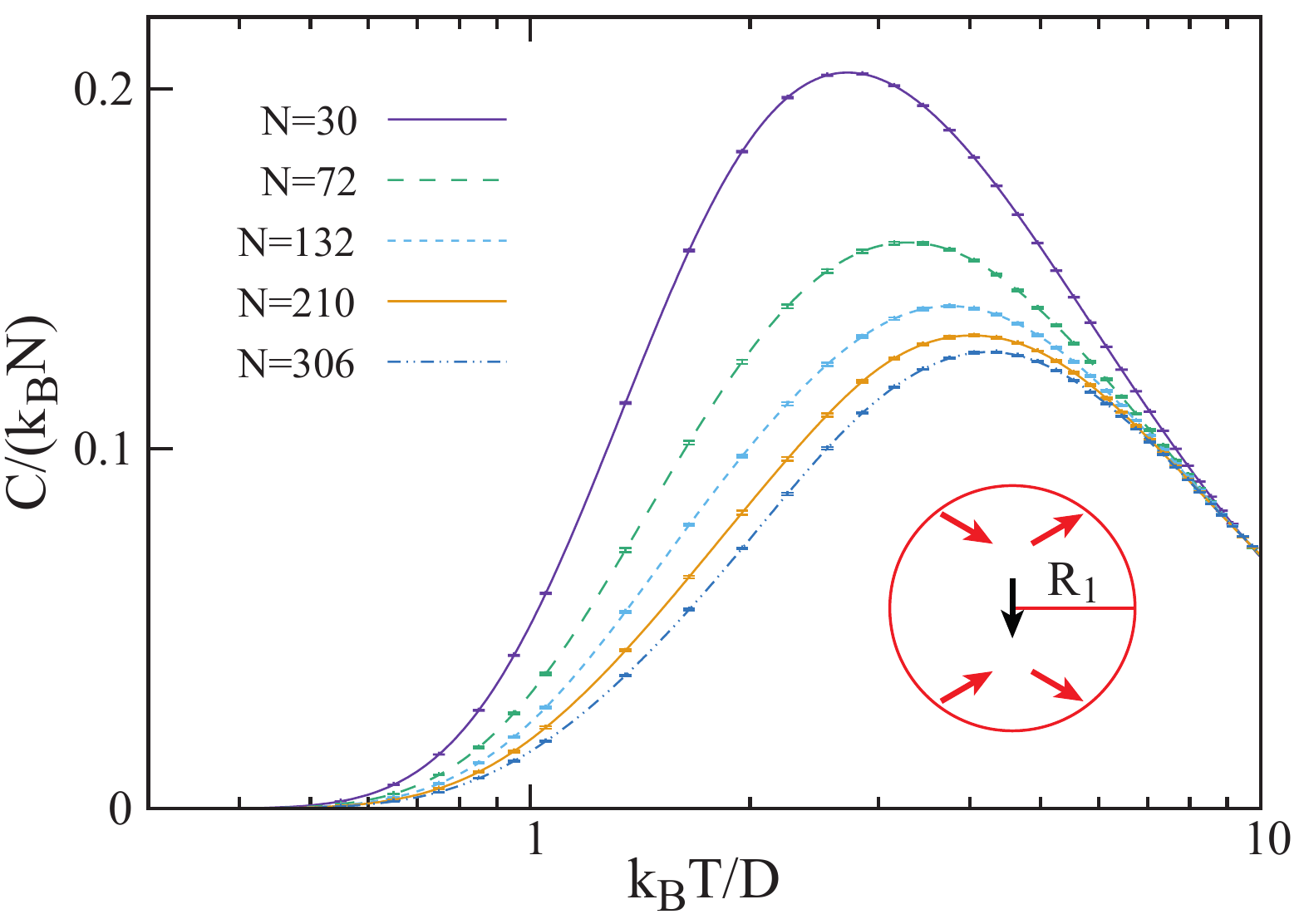}} a) \\
\end{minipage}
\hfill
\begin{minipage}[h]{0.49\linewidth}
	\center{\includegraphics[width=1\linewidth]{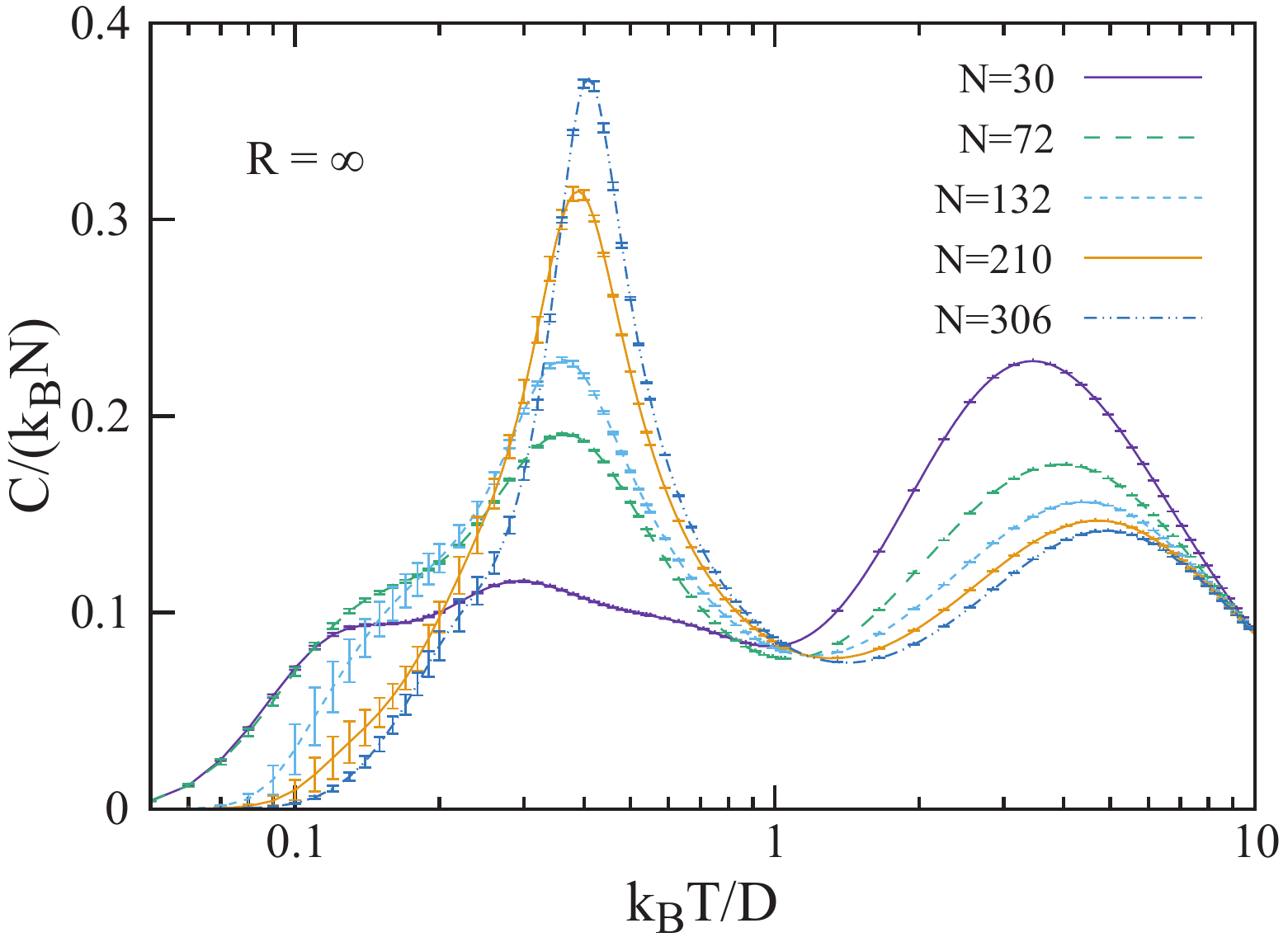}} \\b)
\end{minipage}
\caption{Specific heat as a function of temperature for kagome artificial spin ice (hexagonal sample with $N$ macrospins) with cutoff interaction radius till nearest neighbors (a) and without cutoff interaction radius (b). From \cite{shevchenko2017multicanonical}.}
\label{Hex_LR}
\end{figure}

When one modeling the magnetic systems, one is surely faced with the question of optimizing the code and the used model. Since the dipole-dipole interaction is long-range, it is computationally expensive to simulate the magnetic systems without a cutoff radius, and many researchers cut off the dipole-dipole interaction \cite{silva2012thermodynamics,levis2013thermal, santamaria2000dipolar, rapini2007phase, siddharthan1999ising}. The question naturally becomes, how much can the dipole interaction be cutoff, so that it does not affect the basic thermodynamic characteristics of the model? This is the question that was asked in \cite{shevchenko2017multicanonical}. Rougemaille in his work write, that macrospins in artificial spin ice are coupled via the magnetostatic interaction, which extends beyond nearest neighbors and that including this dipole interaction to the model remove the ground state degeneracy \cite{rougemaille2011artificial}. His work shows the differences between short-range and long-range interaction models. This problem was also considered in \cite{mol2011phase, den2000dipolar}. Also our scientific group in previous works describes this problem in \cite{shevchenko2017multicanonical,shevchenko2017effect}. It was shown \cite{shevchenko2017multicanonical} that the dipole-dipole interaction cutoff to the nearest neighbors ($R_1$ in Fig. \ref{dipole_model}b, includes 4 nearest neighbors) leads to a loss of the low-temperature phase transition (see Fig. \ref{Hex_LR}).

\subsection{Modeling method}
\label{S:3}

In this paper, the Wang-Landau method, one of the modern Monte Carlo methods was used. A detailed description of the method can be found both in the original work by Wang \cite{wang2001efficient}, and in our previous work \cite{shevchenko2017multicanonical}.

It should be noted that 24 cycles of WL sampling were performed. For each calculation, we took an average of 10 samples and calculated the corrected sample standard deviation.  For the systems without dilution, this means that
we performed simulations with different random-number sequences. For a diluted system, this means that we performed simulations with different random configurations of dilution.  

\section{Results and discussions}
\label{S:4}

\subsection{Dipole-dipole interaction radius limitation}

In \cite{shevchenko2017multicanonical} two types of samples (square and hexagonal forms) which are consisting of various numbers of particles formed in artificial kagome spin ice arrays with the different radius of dipole-dipole interaction was simulated by multicanonical Wang-Landau method. Models was considered with ``long-range'' (when each dipole interacts with each other without cutoff radius, $R=\infty$) and ``short-range'' (when each dipole interacts only with the nearest neighbors, the red spins inside the radius $R_1$ in Fig. \ref{dipole_model}b) dipole-dipole interactions. The temperature dependence of the specific heat in ``long-range'' dipole interactions models shows anomalous character, there are two temperature peaks, i.e., critical behavior of the system of dipoles in the region of critical temperature significantly changes. As is known, the evidence of the phase transition presence can be increase of the specific heat peak with increasing size of the system (number of particles in the system). This behavior of specific heat at low temperature can be observed in Fig. \ref{Hex_LR}b. The first peak with increasing number of particles increases and the second peak reduces (Fig. \ref{Hex_LR}b). The specific heat peak of models with ``short-range'' dipole interactions with an increasing number of particles is reduced (Fig. \ref{Hex_LR}a). The exact solution of heat capacity per spin for a small kagome artificial spin ice system (30 particles) is given in Fig. 3b in the paper \cite{shevchenko2017effect}, and it fully coincides with the WL simulation (Fig. \ref{Hex_LR}), which shows the good reliability of the obtained results. 

Thus, the cutoff of dipole-dipole interaction to the nearest neighbors leads to the low-temperature phase transition disappearance. So, the phase transition is provided not by short-range neighbors \cite{shevchenko2017multicanonical}, but what kind of neighbors provide it? 
To investigate this question, it is necessary to increase the dipole-dipole interaction radius and observe the behavior of the low-temperature specific heat peaks.

\begin{figure}
	\begin{minipage}[h]{0.49\linewidth}
		\center{\includegraphics[width=1\linewidth]{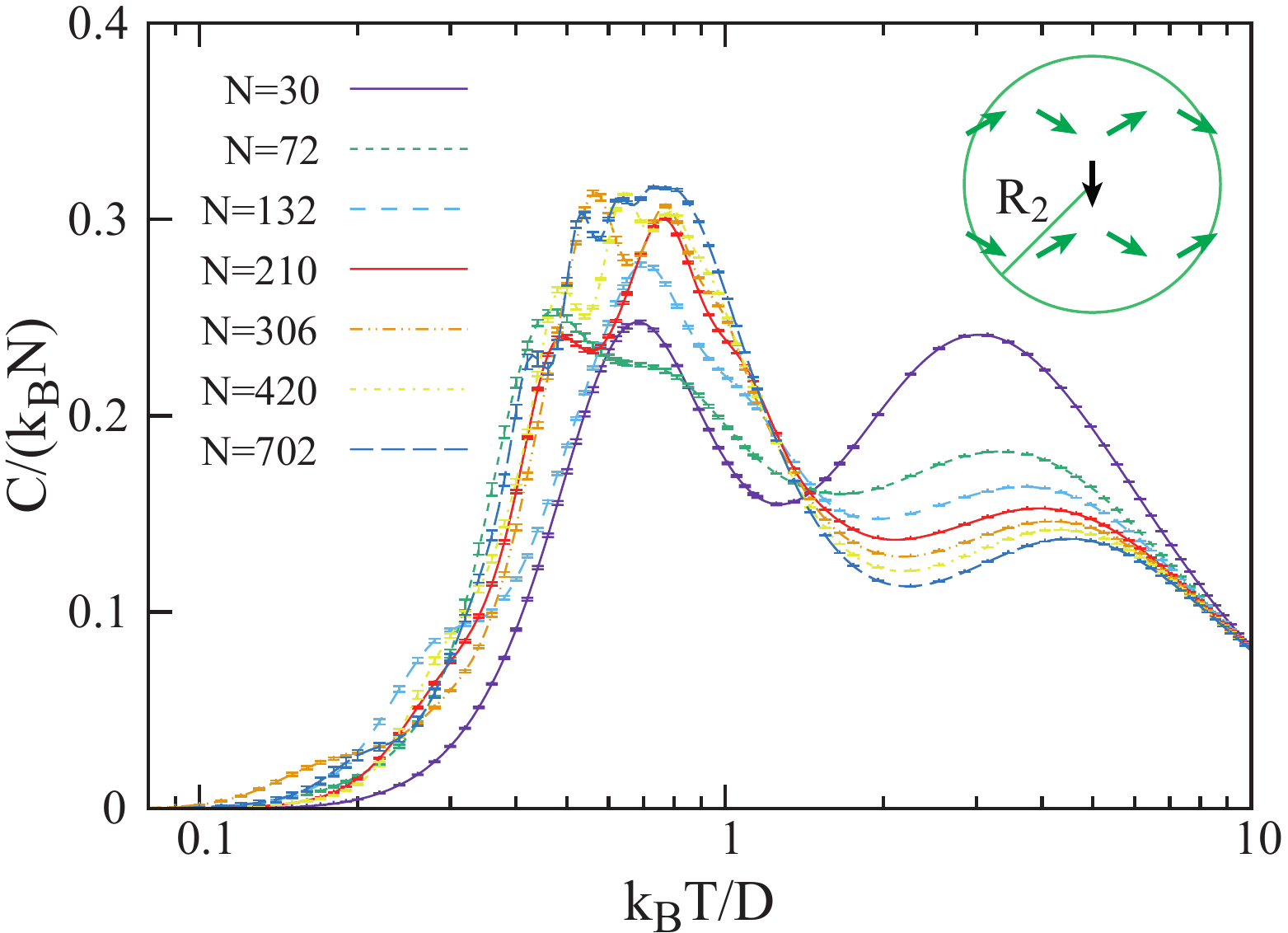}} a) \\
	\end{minipage}
	\hfill
	\begin{minipage}[h]{0.49\linewidth}
		\center{\includegraphics[width=1\linewidth]{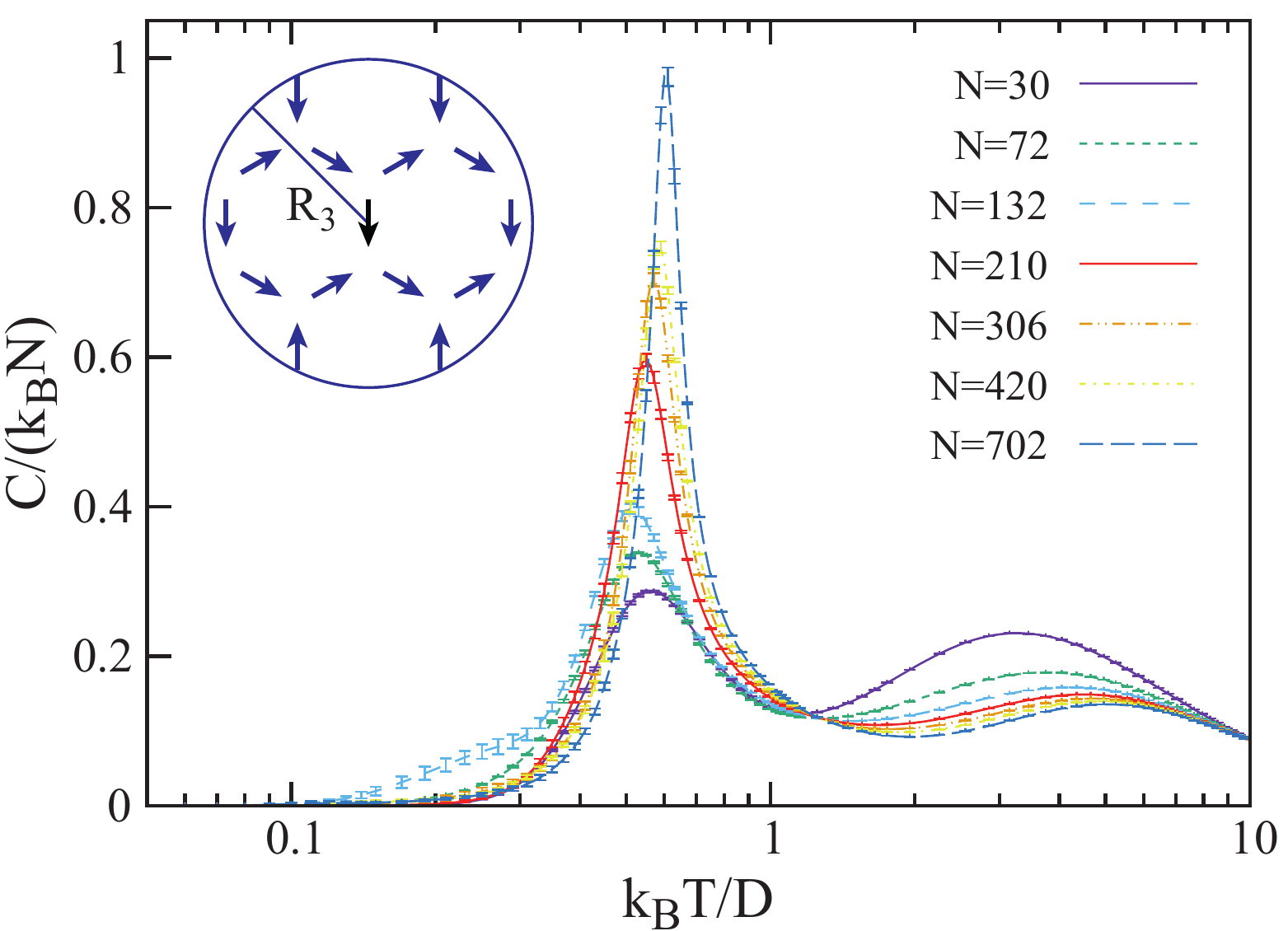}} \\b)
	\end{minipage}
	\caption{Specific heat as a function of temperature for kagome artificial spin ice (hexagonal sample with $N$ macrospins) with cutoff dipole-dipole interaction radius $R_2$, which includes 8 nearest neighbors (a) and $R_3$, which includes 14 nearest neighbors (b).}
	\label{Hex_2_3R}
\end{figure}

As it was shown, the inclusion of only the first coordination sphere $R_1$ (Fig. \ref{dipole_model}b), which includes only 4 nearest neighbors (less at the edges), is insufficient for the phase transition appearance. 

When the next neighbors (total 8 neighboring macrospins inside $R_2$ cutoff radius) are taken into account, a low-temperature peak begins to appear (Fig. \ref{Hex_2_3R}a). However, the low-temperature peak has a very strange behavior, apparently due to the neighbors location anisotropy. Since at $R_2$ cutoff radius the neighbors are located mainly along the perpendicular to the macrospin magnetic moment, they create an uncompensated field on the considering macrospin, which leads to the blurring of the low-temperature peak. In \cite{magomedov2018density} the authors investigated the kagome lattice with short-range exchange interactions, taking into account the first and second neighbors. In this study, the specific heat in the low-temperature region began to split into peaks in a similar way. One can also notice that the high-temperature specific heat peak remains and behaves in a similar way (decreases with increasing system size) as in the model with $R_1$ cutoff radius on Fig. \ref{Hex_LR}.

With a further interaction radius increase to $R_3$, which have 14 nearest neighbors, all the irregularities in the low-temperature specific heat peak disappear (Fig \ref{Hex_2_3R}b). As in the model without cutoff dipole-dipole interaction radius, the low-temperature peak of the specific heat increases, while the high-temperature peak decreases with increasing amounts of macrospins in the system. A further increase of the dipole-dipole interaction radius does not change the behavior of the specific heat, only slightly shifts the peak to the region of lower temperatures, in the limit tending to the phase transition temperature in the model without the interaction radius limiting.

The nature of the specific heat peaks in kagome artificial spin ice is described in detail in the papers \cite{canals2016fragmentation,montaigne2014size}. According to these works, there are four states of kagome artificial spin ice, and they can be clearly understood using dumbbell magnetostatic charge model. At high temperatures, artificial spin ice is in paramagnetic state, where all states are realized. As the temperature decreases, the highest energy states (in charge notation, these are vertices with charge $\pm3$) begin to disappear, and the transition from the paramagnetic state to the ``Spin Ice I'' state occurs. In Spin Ice I state the charge of each lattice vertex is either ``$+$'' or ``$-$'' but they are located chaotically on the lattice, and there are no vertices with charge ``$\pm3$'' left. The high-temperature peak of specific heat (at $\approx$ 3.5 $k_BT/D$) shows the transition of the system from the paramagnetic state to the Spin Ice I state. A further temperature decrease brings the system from Spin Ice I state to the Spin Ice II state, where charges of all vertices are ordered (each vertex with ``$+$'' is adjacent to three vertices ``$-$'' and vice versa). Such ``crystallization'' of  vertex charges on the Kagome lattice gives the phase transition and the low-temperature peak of specific heat (at $\approx$ 0.4 $k_BT/D$). Eventually, the system comes to ``Long-Range Order'' state, where hexagons start to behave like meshing gearwheels.

\subsection{Percolation threshold}

\begin{figure}
	\begin{minipage}[h]{0.49\linewidth}
		\center{\includegraphics[width=0.9\linewidth]{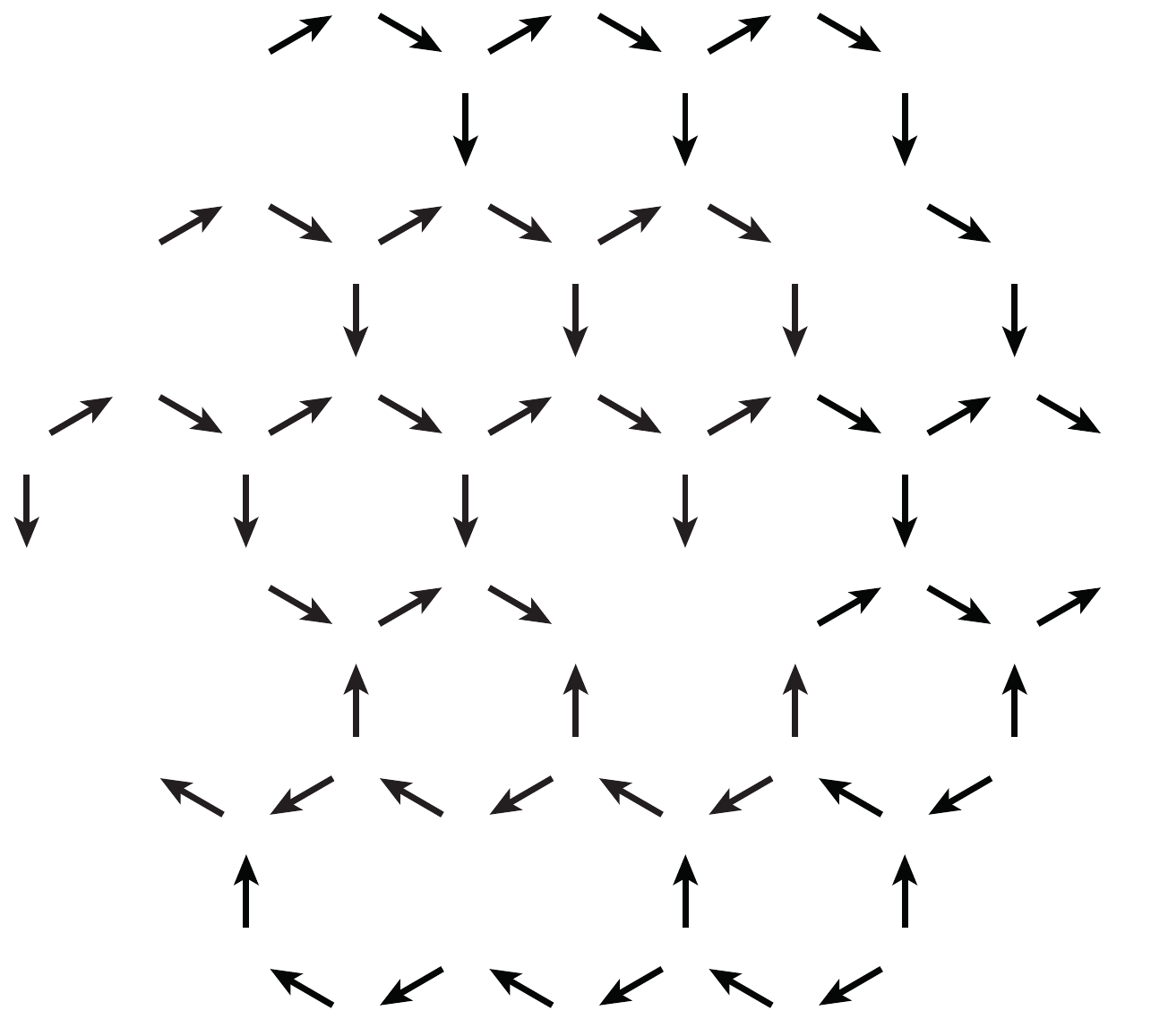}} \\ a)
	\end{minipage}
	\hfill
	\begin{minipage}[h]{0.49\linewidth}
		\center{\includegraphics[width=1\linewidth]{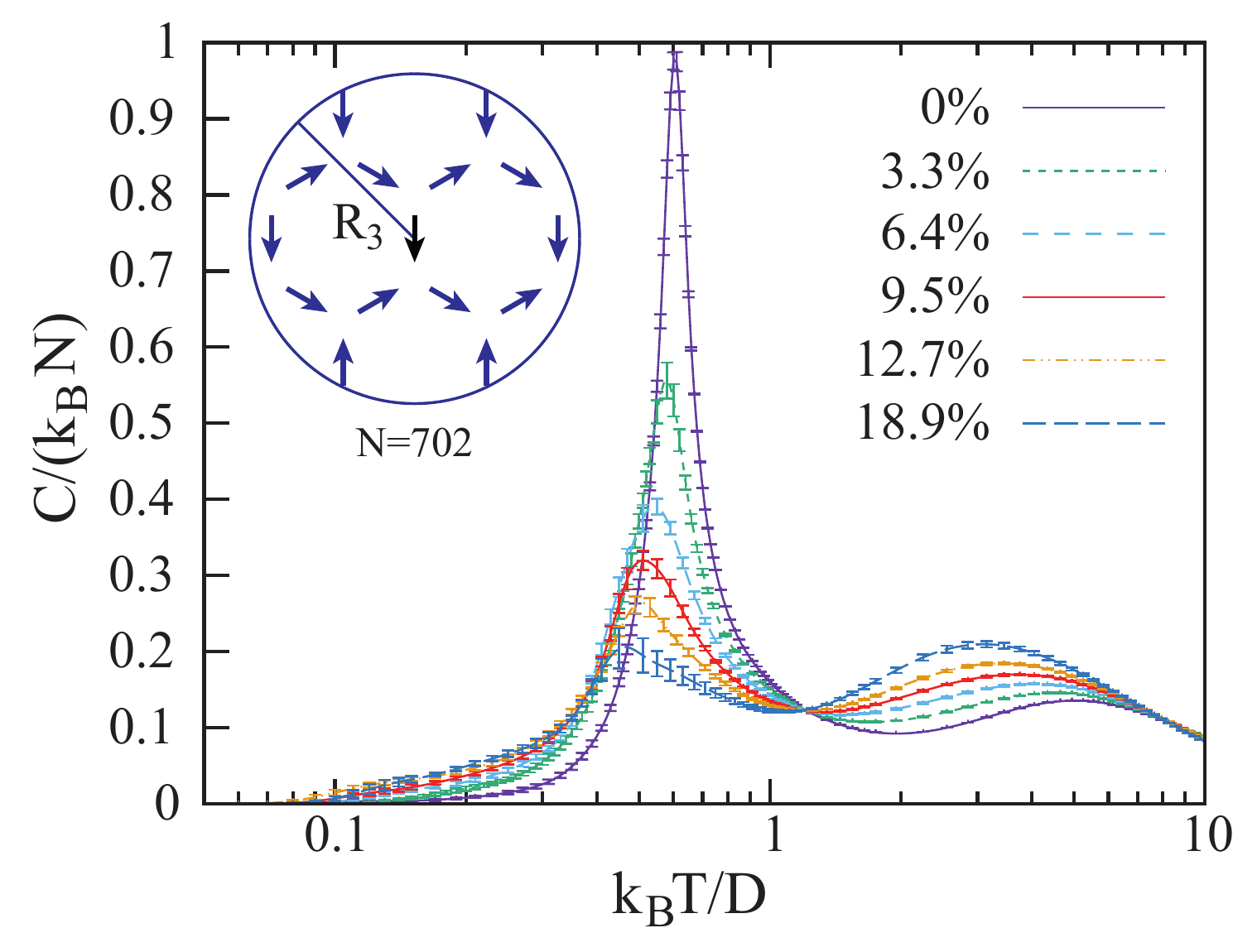}} \\b)
	\end{minipage}
	\caption{a) Example of diluted kagome artificial spin ice, hexagonal sample with $N=72$ macrospins and $13.8\%$ of dilution (random 10 spins was removed). b) Specific heat as a function of temperature for kagome artificial spin ice (hexagonal sample with $N=702$ macrospins) with cutoff dipole-dipole interaction radius $R_3$, which includes 14 nearest neighbors.}
	\label{Diluted}
\end{figure}

In addition to the dipole-dipole interaction radius influence, the effect of dilution on the phase transition existence was also investigated. Dilution occurred by accidentally removing a certain number (corresponding dilution concentration) of macrospins (magnetic islands) from the system. The example of the system with $N=72$ and $13.8\%$ of dilution, which means that 10 spins have been removed is in Fig. \ref{Diluted}a. The temperature dependence of specific heat for kagome artificial spin ice (hexagonal sample with $N=702$ macrospins) with cutoff dipole-dipole interaction radius $R_3$, which includes 14 nearest neighbors and different dilution concentrations is on Fig. \ref{Diluted}b. It can be seen that as the dilution increases, the low-temperature peak of the specific heat decreases rapidly. And at a $20\%$ dilution concentration, the height of the low-temperature peak of specific heat approximately corresponds to the height of the high-temperature peak of specific heat.

\begin{figure}
	\centering\includegraphics[width=0.7\linewidth]{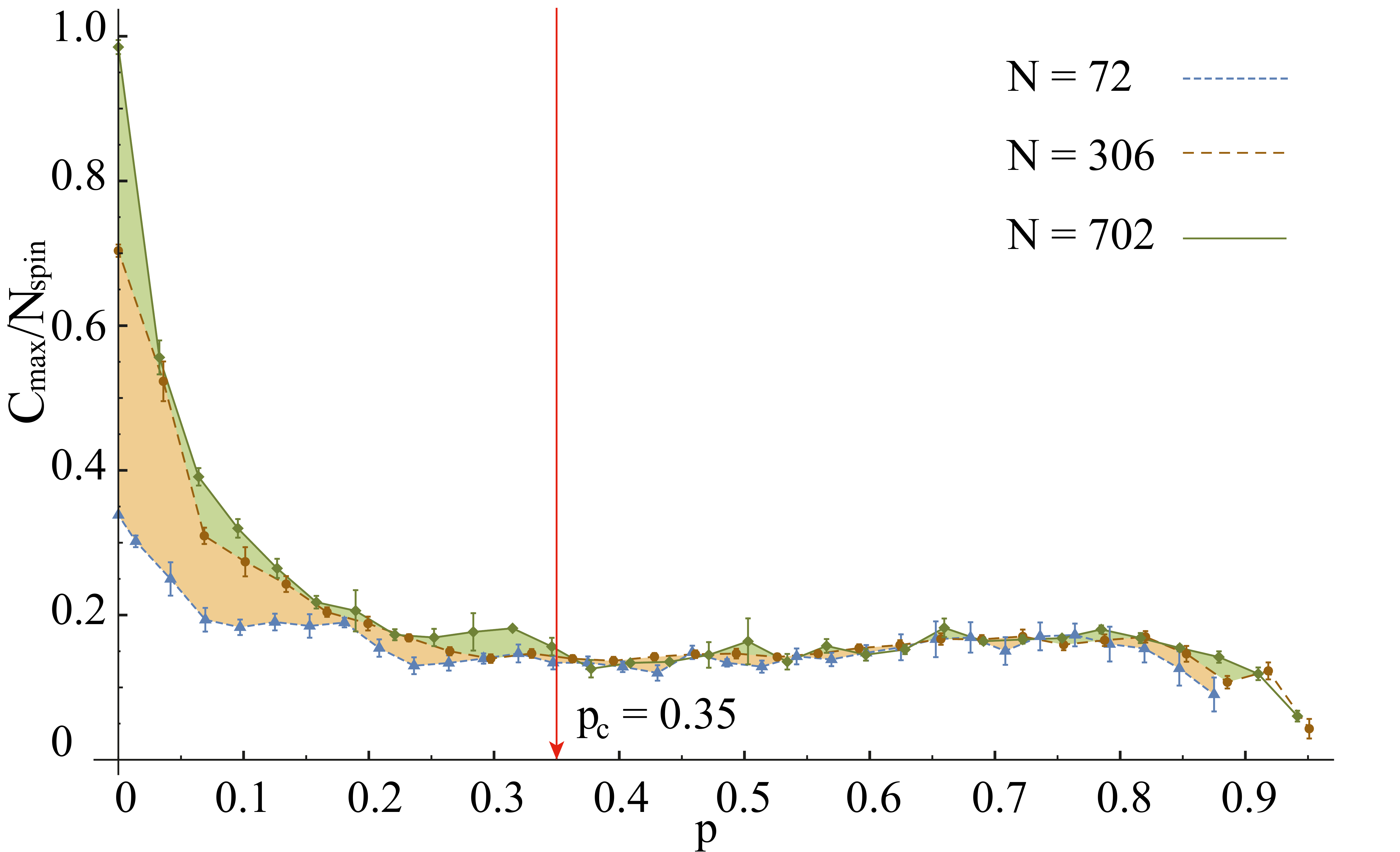}
	\caption{The dependence of the low-temperature peak value of the specific heat of the kagome artificial spin ice model with $R_3$ cutoff dipole interaction radius (which includes 14 nearest neighbors) on the dilution concentration $p$. The vertical red line denotes the theoretical value of the percolation threshold \cite{sykes1964exact}.}
	\label{dilution}
\end{figure}

It was interesting to find the dilution concentration at which the phase transition disappeared. As an indirect parameter showing the phase transition, one can use the increase of the specific heat peak with increasing system size. At critical dilution concentration and higher, the low-temperature specific heat peak will not increase with an increase the system size. In 1964 Sykes and Essam obtained a theoretically bond percolation threshold for the kagome lattice $p_c=2sin(\pi/18)\approx0.35$ \cite{sykes1964exact}. In Fig. \ref{dilution}, the dependence of the low-temperature peak value of the specific heat of kagome artificial spin ice with $R_3$ cutoff interaction radius (14 nearest neighbors) on the dilution concentration is calculated for systems with $N=72,306$ and $702$ macrospins. Each point on the graph corresponds to the averaging of approximately 10 random dilutions for each concentration. The green area shows the concentration where the low-temperature peak value of the specific heat of the systems with $N = 702$ is higher than the systems with $N = 306$. The brown area shows the concentration where the low-temperature peak value of the specific heat of the system with $N = 306$ is higher than the systems with $N = 72$. One can be seen that the specific heat peak growth ceases at the critical dilution concentration $p_c=0.35$ region, and with further dilution, the peak value does not change depending on the system size. That is, the phase transition appears at less than $0.35$ dilution concentration, which gives a good agreement with the Sykes and Essam theory.

\section{Conclusion}

When modeling magnetic systems with long-range interactions, always with the caution should be approached to the interaction radius limiting. As has been shown in many articles, limitation may lead to a fundamental change in the thermodynamic characteristics, for example, the phase transition disappearance. It was shown that in the case of kagome artificial spin ice, at least 3 coordination spheres (with 14 nearest neighbors) should be taken into account. Restriction to a smaller cutoff radius leads to significant changes in the thermodynamic behavior of the main characteristics of the system. An increase of the interaction radius shifts the phase transition temperature to the low-temperature area.

The dilution effect on the phase transition in kagome artificial spin ice was also investigated. It was shown that the phase transition occurs up to $p_c = 0.35$ dilution concentrations, that well coincides with the Sykes and Essam theory. Further dilution leads to the phase transition disappearance.

\section{Acknowledgments}

This work was supported by RFBR according to the research project No. 18-32-00557 (mol\_a) and by the grant \#SP-4348.2018.5 from the President of the Russian Federation for young scientists and graduate students, in accordance with the Program of Development Priority Direction ``Strategic information technologies, including the creation of supercomputers and software development''. I would like to thank Alexey Kurtyshev for the help of data processing and Professor K. Nefedev for his comments.






\bibliography{sample}








\end{document}